\documentclass[twocolumn,showpacs,preprintnumbers,amsmath,amssymb]{revtex4}
\usepackage{graphicx}
\begin{document}

\title{Excitation and decay of projectile-like fragments formed 
in dissipative peripheral collisions at intermediate energies}

\author{R. Yanez}
\author{S. Hudan}
\author{R. Alfaro}
\altaffiliation[]{Present address: Institute of Physics, UNAM, Mexico}
\author{B. Davin}
\author{Y. Larochelle}
\altaffiliation[]{Present address: OakRidge National Laboratory, OakRidge, TN.}
\author{H. Xu}
\altaffiliation[]{Present address: Institute of Modern Physics, CAS, Lanzhou, China.} 
\author{L. Beaulieu}
\altaffiliation[] {Present address: Universit\'e Laval, Quebec, Canada.}
\author{T. Lefort}
{Present address: Universit\'e de Caen, Caen, France.}
\author{V.E. Viola} 
\author{R.T. de Souza} 
\affiliation{
Department of Chemistry and Indiana University Cyclotron Facility \\ 
Indiana University, Bloomington, IN 47405} 

\author{T.X. Liu}
\author{X.D. Liu}
\author{W.G. Lynch}
\author{R. Shomin}
\author{W.P. Tan}
\author{M.B. Tsang} 
\author{A. Vander Molen} 
\author{A.Wagner}
\altaffiliation[] {Present address: Forschungszentrum 
Rossendorf e.V., Dresden, Germany. }
\author{H.F. Xi}
\affiliation{
National Superconducting Cyclotron Laboratory and Department of
Physics and Astronomy \\ 
Michigan State University, East Lansing, MI 48824}

\author{R.J. Charity}
\author{L.G. Sobotka}
\affiliation{
Department of Chemistry, Washington University, St. Louis, MO 63130}

\date{\today}

\begin{abstract}
Projectile-like fragments (PLF:15$\le$Z$\le$46) formed in peripheral and 
mid-peripheral 
collisions of $^{114}$Cd projectiles with $^{92}$Mo nuclei at E/A=50 MeV 
have been detected at very forward angles, 
2.1$^{\circ}$$\le$$\theta_{lab}$$\le$4.2$^{\circ}$. Calorimetric analysis 
of the charged particles observed in coincidence with the PLF reveals that the 
excitation of the primary PLF
is strongly related to its velocity damping.
Furthermore, for a given V$_{PLF*}$, its excitation 
is not related to its size, Z$_{PLF*}$. 
For the largest velocity damping, the excitation energy attained 
is large, 
approximately commensurate with a system at the limiting temperature.

\end{abstract}
\pacs{PACS number(s): 25.70.Mn} 

\maketitle

The nuclear equation-of-state (EOS), 
and in particular its isospin dependence, 
is a topic
of fundamental interest \cite{Li02}. 
For finite nuclei, examination of the limiting temperature, the maximum temperature
attainable \cite{Bonche,Natowitz02a,Natowitz02b},
provides information on the interplay between the nuclear EOS and 
Coulomb instabilities. 
To date, a variety of means has been used  to excite finite 
nuclear matter to the 
limits of stability. These approaches range from
multi-GeV hadronic probes \cite{Lefort01, Beaulieu01} or 
projectile fragmentation at relativistic energies \cite{Scharenberg01,Ogilvie91}, 
to central collisions of intermediate-energy
heavy-ions \cite{Piasecki91,Bowman91,MArie97}. 
Recent advances in the availability of high-intensity 
radioactive beams 
in the intermediate energy domain raises the question of how the N/Z 
dependence of the EOS is best probed.
While it is widely accepted that nuclear matter at high excitation can 
be formed by
central collision of two heavy-ions, 
this approach has several drawbacks in probing the N/Z dependence of the 
EOS.  
Interaction of a projectile with extreme N/Z with a stable target
results in a system with N/Z less exotic than the projectile. 
Furthermore, the fragmenting source 
in a central collision is poorly defined in size, density, and  N/Z.
In contrast to central collisions, peripheral collisions 
result, with large cross-section, in the survival of 
a projectile-like nucleus and a target-like nucleus both at near normal 
density. Previous studies have examined the decay of the excited 
projectile-like fragments produced in peripheral collisions 
\cite{Lott92,Peter95,Dore00}.

In this manuscript, 
we focus on the correlation between the
excitation of the projectile-like fragment (PLF) resulting from peripheral 
and mid-peripheral collisions, 
and its associated velocity damping.
When selected on velocity dissipation, we determine that the size of the PLF
has essentially no impact on its total excitation.  
Moreover, our results indicate that 
mid-peripheral collisions of two intermediate-energy heavy-ions can result 
in formation of a projectile-like fragment excited to the limits of stability.

The experiment was conducted at Michigan State University 
where a beam of $^{114}$Cd nuclei accelerated by the K1200 cyclotron 
to E/A=50 MeV impinged on a $^{92}$Mo foil 5.45 mg/cm$^2$ thick. At 
very forward angles (2.1$^{\circ}$$\le$$\theta_{lab}$$\le$4.2$^{\circ}$) 
charged reaction products were detected 
by an annular ring Si/CsI(Tl) telescope (RD) \cite{davinthesis}. 
For Z$\le$48, this telescope provided both good charge  
($\delta$Z/Z$\sim$0.25) 
and angular 
($\Delta \theta$=0.125$^{\circ}$; 
$\Delta \phi$=22.5$^{\circ}$) resolution.
Light charged particles (LCP: Z$\le$2) and intermediate mass fragments 
(IMF: 3$\le$Z$\le$10) emitted at larger angles 
(7$^{\circ}$$\le$$\theta_{lab}$$\le$58$^{\circ}$) 
were isotopically identified in 
the Large Area Silicon Strip Array, 
LASSA \cite{davin01,wagner01}. 
Charged-particle multiplicity 
was measured using the MSU Miniball \cite{Miniball} 
and Washington 
University Miniwall. A trigger condition of 
three charged particles in the Miniball/Miniwall was used 
during the experiment.

An overall qualitative description of the reaction is provided in Fig.~1. 
With increasing charged-particle multiplicity, N$_c$ the most probable 
atomic number of the projectile-like fragment (PLF), Z$_{PLF}$, 
detected in the RD decreases. While the relationship
between these two quantities for N$_c$$\le$15 is approximately linear, for 
larger multiplicities Z$_{PLF}$ depends more weakly on N$_c$. Moreover, while 
the correlation between the most probable values is clearly evident, it is 
noteworthy that the distribution is broad in both Z$_{PLF}$ and N$_c$.
The relationship between the size of the PLF, as represented by Z$_{PLF}$, 
and the velocity of the PLF, V$_{PLF}$, is shown in Fig.~1b. As the atomic
number of the PLF decreases from Z=48, one observes 
a gradual decrease from the 
projectile velocity, indicated by the arrow i.e. consistent with velocity 
damping of the PLF. For Z$\le$38 the most probable 
velocity is relatively constant with a value of approximately 9.0 cm/ns,
corresponding to 95$\%$ of V$_{beam}$.
In contrast to this constancy, the mimimum velocity attained decreases 
(maximum damping increases) with decreasing Z$_{PLF}$.
While these observations are consistent with previous experimental 
measurements, in this work we focus on the characteristics of the 
PLF: specifically, its 
size (Z$_{PLF}$) and velocity damping.

\begin{figure} 
\includegraphics [scale=0.45]{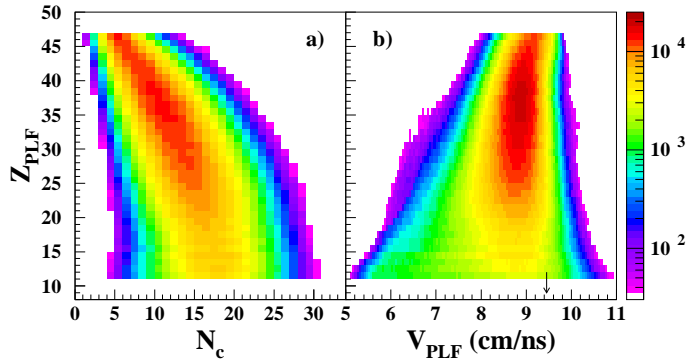}%
\caption{\label{fig:fig0}
(Color online) Panel a) Relationship between the measured charged-particle multiplicity, 
N$_c$, and the atomic number of the detected PLF, Z$_{PLF}$. Panel b) 
Correlation between Z$_{PLF}$ and its velocity in the laboratory, V$_{PLF}$. 
The arrow indicates the beam velocity.} 
\end{figure}

To examine the excitation of the PLF
formed in peripheral collisions, we required the 
detection of a single fragment in the RD with 15$\le$Z$\le$46
and a velocity larger than the 
center-of-mass velocity.
The magnitude of the most probable velocity of this fragment 
(95$\%$ of V$_{beam}$)
suggests that it is a remnant of 
the projectile
following the interaction stage of the collision and de-excitation. 
For clarity, we subsequently refer to the primary excited 
projectile-like fragment 
as the PLF$^*$ and the detected nucleus following de-excitation as the PLF.
The characteristics of the PLF$^*$, namely its Z, A, and velocity 
can be reconstructed by measuring the PLF residue and examining the 
multiplicities, kinetic-energy spectra, and angular distribution 
of coincident particles.

The energy-angle correlation 
for $\alpha$ particles 
detected in LASSA in the reference frame of the PLF$^*$
is displayed in Fig. 2a. 
All quantities have been corrected for the geometric 
efficiency of LASSA, as well as recoil due to emission. 
A prominent feature evident is the
horizontal ridge observed for 
$\theta_{PLF*}$$>$40$^{\circ}$. This 
ridge can be understood as statistical emission of $\alpha$ particles 
from the PLF$^*$. Emission
at small angles $\theta_{PLF*}$$<$30$^{\circ}$ and large angles
$\theta_{PLF*}$$>$160$^{\circ}$ is not observed due to the finite 
experimental acceptance.
At backward angles, $\theta_{PLF*}$$>$120$^{\circ}$, one
observes a second component in the two-dimensional spectrum. 
This component, which exhibits higher average energies, is a general feature 
of non-central intermediate energy heavy-ion collisions and has  
been associated with dynamical processes \cite{Toke95,Bocage00,Davin02}. 
It is evident from Fig.~2a, that for
$\theta_{PLF*}$ $<$ 90$^{\circ}$ the statistical emission process dominates,
suggesting minimal dynamical contamination in this region. 

In Fig. 2b, the $\alpha$ particle kinetic-energy 
spectra are shown for six discrete angles 
45$^{\circ}$$\le$$\theta_{PLF*}$$\le$70$^{\circ}$ with
$\Delta \theta_{PLF*}$=1$^\circ$. To facilitate the
comparison of the different spectra, the 
solid line depicts a single fit, performed for 
$\theta_{PLF*}$ = 55$^{\circ}$, to a Maxwell-Boltzmann
distribution (as described below). 
Enhanced yield in the kinetic-energy 
spectra for E$\ge$40 MeV can be understood as prompt 
$\alpha$ emission prior to attainment of equilibrium by the PLF$^*$.
As the exponential slopes of the kinetic-energy spectra 
reflect the excitation 
of the emitting source \cite{benenson94}, 
the overlap of the different spectra with the single exponential 
in Fig.~2b 
indicates that in this angular range, both the $\alpha$ yield 
and emission temperature are relatively constant, consistent with 
decay of a single isotropically emitting source.

\begin{figure} 
\includegraphics [scale=0.45]{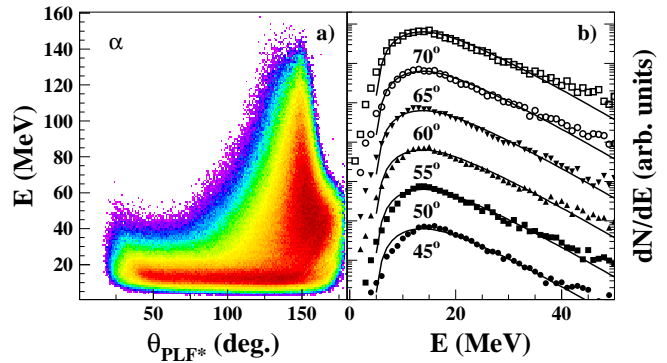}%
\caption{\label{fig:fig1}
(Color online) Panel a) Energy-angle correlation 
for $\alpha$ particles, in the 
PLF$^*$ frame. Color respresents the 
relative probability on a logarithmic scale.
Panel b) Kinetic-energy spectra for $\alpha$ particles in the
PLF$^*$ frame for the indicated angles with a 
vertical displacement of a factor of 10 for clarity.
} 
\end{figure}

\begin{figure} 
\includegraphics [scale=0.4]{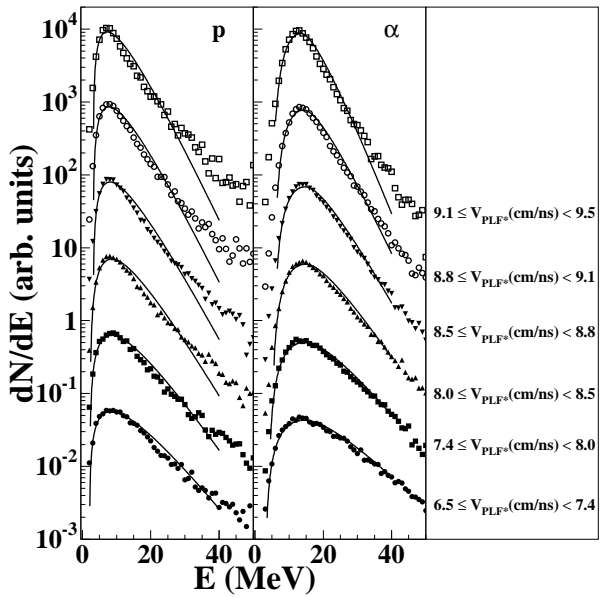}%
\caption{\label{fig:fig2}
Kinetic-energy spectra for p and $\alpha$ particles emitted
forward of the PLF$^*$ associated with different velocity, V$_{PLF*}$. 
Velocity selections are 
shown in the right hand panel. 
Spectra
have been displaced vertically by a factor of 10 for clarity.
Solid lines depict the result of 
fitting the spectra with a Maxwell-Boltzmann distribution. See text 
for details.
}
\end{figure}

The correlation between the excitation of the PLF*, evidenced by the kinetic 
character of emitted particles, and  
its velocity damping is examined in Fig.~3. 
In this figure, the kinetic-energy distributions of 
p and $\alpha$, 
emitted in the range 
40$^{\circ}$$<$$\theta_{PLF*}$$<$75$^{\circ}$, 
are shown selected on 
decreasing velocity of the PLF*, V$_{PLF*}$. 
One qualitatively observes that all the distributions are approximately 
Maxwellian in shape with a notable flattening of the exponential slope 
with decreasing V$_{PLF*}$ i.e. increased velocity damping. For the 
uppermost spectra, which correspond to the lowest velocity damping, 
deviation from the single 
exponential behavior is evident. This deviation from the single exponential 
behavior, which does
not contribute significantly to the total yield of detected particles
can be attributed to pre-equilibrium emission. At low velocity damping,
i.e. low excitation, suppression of the statistical component maximizes 
the separation of the pre-equilibrium and equilibrium components. Comparison 
of the measured yield with the single source fit suggests 
that these pre-equilibrium processes constitute $\approx$2$\%$ of the total 
PLF$^*$ yield.
To quantify the dependence of the exponential spectral tail on
V$_{PLF*}$ for isotopically identified particles, 
we have fit the spectra selected on V$_{PLF*}$. 

The fitting function used was \cite{Lestone}:

\begin{mathletters}
\begin{equation}
N(\epsilon)=0, \text{  if  } \epsilon \leq B^\prime,
\end{equation}
\begin{equation}
N(\epsilon) \propto C^\prime \left(\epsilon-B^\prime \right)^D \exp \left(-\frac \epsilon T_{s} \right), \text{  if  } B^\prime < \epsilon < B+T_{s},
\end{equation}
\begin{equation}
N(\epsilon) \propto \left(\epsilon-B \right) \exp \left(-\frac \epsilon T_{s} \right), \text{  if  } \epsilon \geq B+T_{s},
\end{equation}
\end{mathletters}

\noindent
where $C^\prime=T_{s}/\left(DT_{s}\right)^D$ and $B^\prime=\left(1-D\right)T_{s}+B$.

\noindent The parameter $\epsilon$ represents the kinetic energy, while
B is a barrier parameter
and D a barrier diffuseness and penetrability parameter.
The parameter T$_{s}$ which characterizes the exponential can be 
related within a statistical framework to the energy dependence of the 
density of states in the daughter nucleus following decay. The extracted
slope parameter, T$_{s}$ reflects the average excitation of the 
distribution of
nuclei decaying {\it via} the specific particle decay selected.
The resulting fits are shown as solid lines in Figs. 2 and 3. 
A qualitative feature readily apparent in Fig.~3 is that 
the p and $\alpha$
spectra exhibit exponential tails of similar magnitude for a 
given V$_{PLF*}$. Although similar in magnitude, the extracted slope parameters
are not identical. While at the lowest excitation the slope parameters are
approximately equal, at high excitation the proton slope parameters are 
lower than the $\alpha$ slope parameters by 1.2-1.3 MeV, probably due to 
differences in the de-excitation cascade.

\begin{figure} 
\includegraphics [scale=0.4]{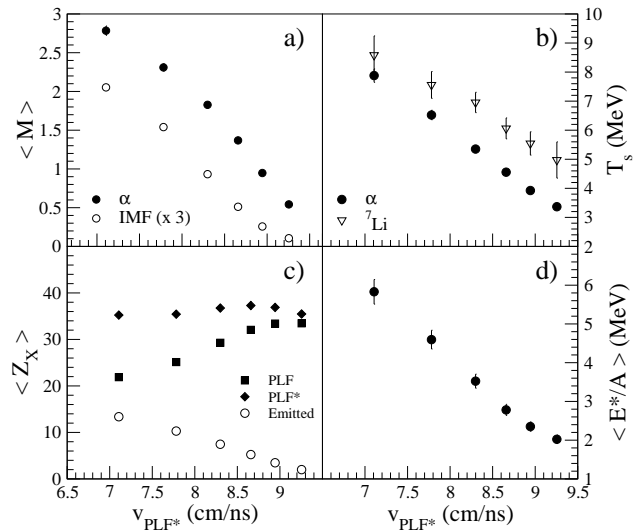}%
\caption{\label{fig:fig3}
Panel a: Extracted average multiplicities for $\alpha$ particles and IMFs
as a function of V$_{PLF*}$. 
Panel b: Relation between the
extracted slope parameters for $\alpha$ particles and $^7$Li fragments emitted
from the PLF$^*$ and V$_{PLF*}$. 
Panel c: Dependence of 
$<$Z$_{emitted}$$>$, $<$Z$_{PLF}$$>$,
and  $<$Z$_{PLF*}$$>$ on V$_{PLF*}$.
Panel d: Excitation energy scale deduced by calorimetry for different 
velocity dissipation. 
}
\end{figure}

To extract the multiplicities of particles emitted from the PLF$^*$, we 
assume isotropic emission from the PLF$^*$, 
consistent with the emission pattern shown in Fig~2. 
This assumption is supported by the
angular distribution of particles forward of the PLF$^*$. For example, for
$\alpha$ particles with E$\le$25 MeV in the PLF$^*$ frame, the yield 
is essentially constant, increasing by $\approx$15$\%$ as 
$\theta_{PLF*}$ increases from 50$^{\circ}$ to 90$^{\circ}$.
By assuming all 
particles emitted forward of the PLF$^*$ (V$_{\parallel}$$>$V$_{PLF*}$) can be attributed to 
the PLF$^*$ and accounting for the geometric acceptance, we 
deduced the total multiplicity of 
charged particles evaporated from the PLF$^*$.
The dependence of the average extracted 
multiplicities, $\langle$M$\rangle$, and slope parameters, 
T$_{s}$, on V$_{PLF*}$ is displayed in Fig.~4a and 4b. 
With decreasing V$_{PLF*}$, 
the average multiplicity of emitted $\alpha$ particles, 
indicated by the filled symbols in Fig.~4a, increases 
from 0.5 to 2.75, a factor of approximately 5. The 
multiplicity of IMFs (3$\le$Z$\le$8) exhibits an
increase from $\approx$0.03 to $\approx$0.7 over the same interval. 
Along with the increase in the charged-particle multiplicity, the 
slope parameter T$_{s}$ 
for $\alpha$ particles increases from 3.4 to 8.0, 
i.e. by a factor of approximately 2 as shown by the filled circles
in Fig.~4b. The dependence
of T$_{s}$
on V$_{PLF*}$ for $^7$Li fragments -- representative of IMFs -- is 
also shown in Fig.~4b by the open symbols.
The monotonic increase with velocity damping observed for $\alpha$ 
particles is also manifested for $^7$Li fragments. 
In comparison to the $\alpha$ slopes, the T$_{s}$ values for $^7$Li fragments 
are larger in magnitude, most likely reflecting an
earlier emission time as compared to $\alpha$ particles.
The monotonic increase in both T$_{s}$ 
and $\langle$M$\rangle$ 
with decreasing V$_{PLF*}$ can be understood as 
the increased excitation of the PLF$^*$ with increasing 
velocity dissipation. 
From the measured multiplicities of emitted particles, 
we have calculated the average emitted charge, from the PLF$^*$ 
$\langle$Z$_{emitted}$$\rangle$ = $\sum$$\langle$M$_i$$\rangle$Z$_i$.
With increasing velocity dissipation, 
$\langle$Z$_{emitted}$$\rangle$ increases from $\approx$2 to
$\approx$14 as shown in Fig.~4c by the open symbols. 
Coupled with this increase is 
a concurrent decrease of the detected 
$\langle$Z$_{PLF}$$\rangle$ from $\approx$34 to $\approx$22, indicated by the
filled squares. As can be seen from the solid diamonds in Fig.~4c, 
$\langle$Z$_{PLF*}$$\rangle$ =
$\langle$Z$_{PLF}$$\rangle$ + $\langle$Z$_{emitted}$$\rangle$ 
has a near constant value of $\approx$36-38, $\approx$75-80$\%$ of Z$_{projectile}$, 
for all values of V$_{PLF*}$. The reason $\langle$Z$_{PLF*}$$\rangle$ 
does not
equal Z$_{projectile}$ even for the lowest velocity dissipation is 
due to the experimental hardware trigger of N$_c$$\ge$3 in
the Miniball/Miniwall, which suppresses the most peripheral collisions. The 
impact of charged-particle multiplicity selection in suppressing the most
peripheral collisions 
has been established previously \cite{Schroder,Kunde}.

In order to determine the excitation of the PLF$^*$ 
we have performed a calorimetric analysis \cite{Steckmeyer89,Cussol93}.  
Using the extracted average multiplicity 
and the measured
average kinetic energy for each type of particle evaporated 
from the PLF$^*$, 
we deduce the average excitation energy of the PLF$^*$ following 
pre-equilibrium emission. 
As the mass of the PLF was not measured, we assumed
its mass based upon the EPAX systematics \cite{Summerer00}.
The average mass of the PLF$^*$ was assumed to be 
given by $\langle$A$_{PLF*}$$\rangle$ = 
(A/Z)$_{projectile}$$\langle$Z$_{PLF*}$$\rangle$. The average neutron 
multiplicity was deduced by mass conservation and their average 
energy of was taken as
$\langle$E$^n$$\rangle$=$\langle$E$^p$$\rangle$-(0.106*Z$_{PLF*}$ - 0.9) 
\cite{Parker91}.
In order to calculate the $\langle$Q$\rangle$
we assumed:
$\langle$Q$\rangle$=$\Delta$m(A$_{PLF*}$,Z$_{PLF*}$)-
($\Delta$m(A$_{PLF},$Z$_{PLF}$)
+$\langle$Q$_{emitted}$$\rangle$)
where 
$\langle$Q$_{emitted}$$\rangle$ = $\sum$$\langle$M$_i$$\rangle$$\Delta$m$_i$
with {\it i} ranging over all charged particles and neutrons emitted, 
where $\Delta$m is the 
mass defect. 
Within this framework, we deduce the relation between 
V$_{PLF*}$ and E$^*$/A shown in Fig.~4d. 

The deduced relationship between E$^*$/A and velocity damping 
exhibits an essentially linear behavior over the entire range of
velocity dissipation.  
This relationship is significant as it 
has been generally accepted that the excitation
of the PLF$^*$ depends on the overlap between projectile and 
target \cite{Peter95} 
and {\it not} 
on velocity dissipation.  
Furthermore, we observe that for the most damped collisions, which 
correspond to $\approx$75$\%$ of V$_{beam}$ 
i.e. considerable velocity damping, an 
excitation of approximately 6 MeV/A is attained. 

It is important to place the maximum excitation observed for these
mid-peripheral collisions in context. The excitation energy observed
is comparable to the excitation for which one observes the onset of 
a plateau in the caloric curve for an A=90 nucleus \cite{Natowitz02b}. 
Moreover, the selection of the survival of a large residue (the PLF) 
favors lower excitation. 
From this analysis it is clear that an excitation comparable to that
observed for central collision of two intermediate energy heavy-ions 
has been attained for
mid-peripheral heavy-ion collisions at E/A=50 MeV. 

To assess the sensitivity of the deduced E$^*$/A 
within the calorimetric approach to the spin of the 
emitting PLF$^*$, we utilized the
statistical model code GEMINI \cite{Charity01}. 
The code GEMINI employs a Hauser-Feshbach formalism to describe 
the statistical emission 
of particles from a nucleus of (Z,A) characterized by an excitation 
energy E$^*$ and a spin J. 
We varied the excitation of the source between 2 MeV/A and 5 MeV/A
with an initial Z = 38 (deduced from the data) and A = 90. 
We explored the sensitivity of the deduced E$^*$/A to spin 
by calculations at J=0 and 40$\hbar$. The results of these calculations
are depicted by the different lines in Fig. ~5.
The calculated $<$Z$_{emitted}$$>$ is quite insensitive to the 
assumed spin of the decaying nucleus for a constant total E$^*$ while 
$\langle$M$_{\alpha}$$\rangle$ manifests a modest sensitivity. 
From the GEMINI calculations we conclude that 
$<$Z$_{emitted}$$>$ is fairly insensitive to the spin
of the decaying nucleus, supporting the excitation scale deduced for the 
data.

\begin{figure} 
\includegraphics [scale=0.4]{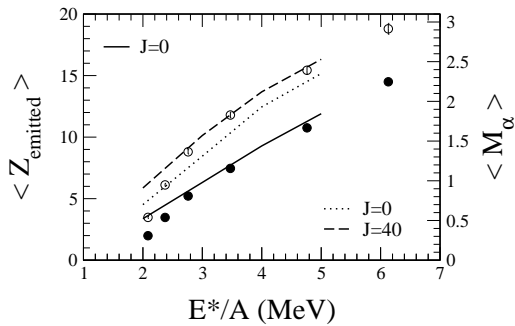}%
\caption{\label{fig:fig4}
Dependence of $\langle$Z$_{emitted}$$\rangle$ and 
$\langle$M$_{\alpha}$$\rangle$
on E$^*$/A. Solid symbols and line refer to $\langle$Z$_{emitted}$$\rangle$ 
while open symbols and broken lines refer to $\langle$M$_{\alpha}$$\rangle$. 
For the experimental data (symbols) the excitation scale is deduced by 
calorimetry while for the GEMINI calculations (lines) 
it corresponds to the initial
excitation of the emitting source. In GEMINI, $\langle$Z$_{emitted}$$\rangle$
for the J=0 and 40 $\hbar$ calculations overlap each other.
}
\end{figure}

We have examined whether the correlation between excitation and velocity
damping observed in Fig.~4d depends on the size of the 
observed PLF, Z$_{PLF}$. The dependence of Z$_{emitted}$, T$_{s}$ , 
and deduced $\langle$E$^*$$\rangle$ on V$_{PLF*}$ is displayed in Fig.~6
selected on Z$_{PLF}$. 
Quite strikingly, one finds 
that the monotonic increase of Z$_{emitted}$ and T$_{s}$ 
with decreasing V$_{PLF*}$, 
previously observed, 
is independent of Z$_{PLF}$. 
Moreover, different Z$_{PLF}$ with the same velocity damping manifest the same 
average total excitation, $\langle$E$^*$$\rangle$, 
resulting in different $\langle$E$^*$/A$\rangle$.
We have 
associated Z$_{PLF*}$ with b/b$_{max}$ in a geometrical model of two 
overlapping nuclei. 
From this geometrical model and the results in Fig.~6, one 
concludes that for 0.4$\le$b/b$_{max}$$\le$1.0, 
velocity dissipation determines the excitation of the PLF*. 
For the impact parameter range examined, the most probable velocity 
dissipation is low, and does not change significantly with 
decreasing Z$_{PLF}$, (Fig.~1b) suggesting that the most probable 
excitation of the projectile 
(and target) spectator is moderate ($\approx$200 MeV) and 
approximately constant with impact parameter. Systems formed at high 
excitation represent a small fraction of the cross-section.

\begin{figure} 
\includegraphics [scale=0.4]{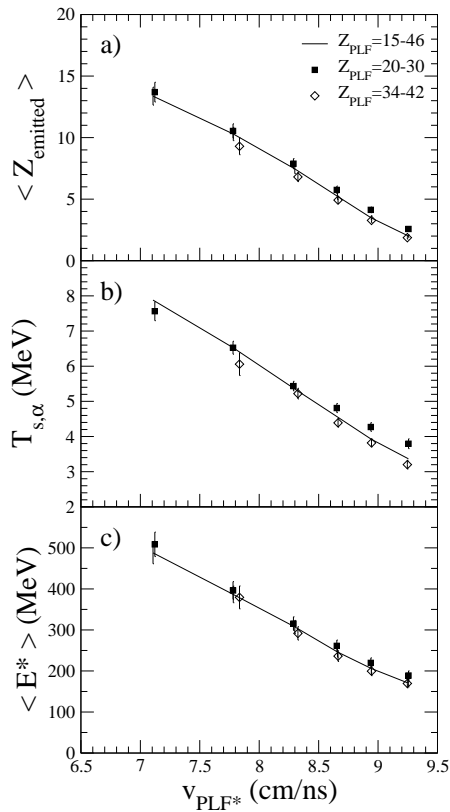}%
\caption{\label{fig:fig5}
Dependence of $\langle$Z$_{emitted}$$\rangle$, T$_{s,\alpha}$ , and $\langle$E$^*$$\rangle$ 
on V$_{PLF*}$ selected on
Z$_{PLF}$.}
\end{figure}

We have carefully examined 
all possible sources of bias in our measurement and analysis and find no 
systematic bias that is responsible for the observed results. 
We therefore turn to understanding the observed correlations.
One possibility in explaining 
these results is that for each impact parameter 
a distribution of contact times exists. While the impact parameter determines 
the size of the PLF$^*$, it is the contact time that determines the velocity
dissipation and the excitation of the PLF$^*$.  
Reaction models in the intermediate energy domain need to account for
the underlying mechanism responsible for the distribution of velocity 
damping (contact time), as well as the fundamental association 
between velocity damping and the PLF$^*$ excitation. 

In summary, for peripheral collisions 
of a near symmetric system at E/A = 50 MeV, 
highly excited projectile-like fragments are formed 
and {\it the excitation of the PLF$^*$ is associated with its velocity dissipation}.
Moreover, when selected on velocity dissipation, {\it the total excitation of the PLF$^*$ does not depend on 
its size}.  
Selection of PLF of different size but the same velocity dissipation 
corresponds to the same excitation energy.
Remarkably, {\it for the largest velocity dissipation an excitation energy 
approximately consistent with the limiting temperature is deduced}
presenting interesting opportunities for probing the N/Z dependence of the
limiting temperature.

\begin{acknowledgments}
	We would like to acknowledge the 
valuable assistance of the staff at MSU-NSCL for
providing the high quality beams which made this experiment possible. 
This work was supported by the
U.S. Department of Energy under DE-FG02-92ER40714 (IU), 
DE-FG02-87ER-40316 (WU) and the
National Science Foundation under Grant No. PHY-95-28844 (MSU).\par
\end{acknowledgments}

\bibliography{ricardo2full.bib} 

\end{document}